\documentclass[preprint]{iucr}              

     \journalcode{J}              
\usepackage{amsmath, amsthm, amssymb, mathrsfs}
\usepackage{graphicx}
\usepackage{relsize}
\usepackage[margin=1.0in]{geometry}
\usepackage{color}
\usepackage{multirow}
\usepackage{dcolumn}

\graphicspath{{./figs/}}

\def\TN{$T_\mathrm{N}$}

\def\muB{$\mu_{\mathrm{B}}$}
\def\deltampdf{3D-$\Delta$mPDF}
\def\diffpympdf{\texttt{diffpy.mpdf}}

\begin{document}                  



\title{DIFFPY.MPDF: Open-source software for magnetic pair distribution function analysis}
\shorttitle{Software for magnetic PDF analysis}


\cauthor[a]{Benjamin~A.}{Frandsen}{benfrandsen@byu.edu}{}
\author[a]{Parker~K.}{Hamilton}
\author[a]{Jacob~A.}{Christensen}
\author[a]{Eric}{Stubben}
\author[b,c]{Simon~J.~L.}{Billinge}

\aff[a]{Department of Physics and Astronomy, Brigham Young University, Provo, UT 84602, \country{U.S.A.}}
\aff[b]{Department of Applied Physics and Applied Mathematics, Columbia University, New York, NY 10027, \country{USA}}
\aff[c]{Condensed Matter Physics and Materials Science Department, Brookhaven National Laboratory, Upton, NY 11973, \country{USA}}









\maketitle                        

\begin{synopsis}
We introduce the open-source software package \texttt{diffpy.mpdf} for magnetic pair distribution function analysis. Written in python and part of the DiffPy suite for diffraction and pair distribution function analysis, this package allows users to build magnetic structures, calculate one- and three-dimensional magnetic pair distribution function patterns, and perform fits to magnetic pair distribution function data.
\end{synopsis}

\begin{abstract}
The open-source python package \texttt{diffpy.mpdf}, part of the DiffPy suite for diffraction and pair distribution function analysis, provides a user-friendly approach for performing magnetic pair distribution function (mPDF) analysis. The package builds on existing libraries in the DiffPy suite to allow users to create models of magnetic structures and calculate corresponding one-dimensional and three-dimensional mPDF patterns. \texttt{diffpy.mpdf} can be used to perform fits to mPDF data either in isolation or in combination with atomic PDF data for joint refinements of the atomic and magnetic structure. Examples are given using MnO and MnTe as representative antiferromagnetic compounds and MnSb as a representative ferromagnet.
\end{abstract}


\section{Introduction}

Magnetic pair distribution function (mPDF) analysis of neutron total scattering data was introduced in 2014 as a method for studying local magnetic correlations in materials~\cite{frand;aca14}. Analogous to the more familiar atomic pair distribution function (PDF) method~\cite{egami;b;utbp12}, mPDF analysis utilizes the Fourier transform of the magnetic neutron scattering cross section to yield information about pairwise magnetic correlations in real space. The method can be used for both powder samples and single crystals~\cite{roth;iucrj18}, the latter case yielding the three-dimensional difference magnetic pair distribution function (\deltampdf). Magnetic PDF analysis has been successfully applied to the study of long- and short-range magnetic correlations in numerous magnetic systems, including strongly correlated electron systems~\cite{frand;aca15,frand;prl16,frand;prb20}, geometrically frustrated magnets~\cite{frand;prm17,roth;prb19,lefra;prb19,frand;prb21}, spin glass systems~\cite{zhang;prb19}, technologically relevant ferromagnets and antiferromagnets~\cite{frand;prb16,kodam;jpsj17, baral;matter22}, and others~\cite{tripa;prb19}. Here, we present \texttt{diffpy.mpdf}, the first full-featured, open-access software package for mPDF analysis.

\section{Theoretical Background}
\subsection{One-dimensional mPDF}
For an assembly of localized spins belonging to a single species of magnetic atom within an isotropic powder sample, the standard one-dimensional (1D) mPDF is given by~\cite{frand;aca14, kodam;jpsj17}
\begin{align}
G_{\mathrm{mag}}(r)&=\frac{2}{\pi}\int_{Q_{\mathrm{min}}}^{\infty} Q\left(\frac{\left(\text{d}\sigma/\text{d}\Omega\right)_{\mathrm{mag}}}{\frac{2}{3}N_sS(S+1)(\gamma r_0)^2 [f(Q)]^2}-1\right)\sin{(Q r)} \text{d}Q \label{FT}
\\&=\frac{3}{2 S(S+1)}\left(\frac{1}{N_s}\sum\limits_{i \ne j}\left[ \frac{A_{ij}}{r}\delta (r-r_{ij})+B_{ij}\frac{r}{r_{ij}^3}\Theta (r_{ij}-r)\right] - 4\pi r \rho_0 \frac{2}{3} m^2\right) \label{fullfofr},
\end{align}
where the first equation represents the experimental definition of the mPDF and the second equation is the theoretical mPDF for a given magnetic structure. In these equations, $r$ is real-space distance, $Q$ is the magnitude of the scattering vector, $Q_{\mathrm{min}}$ is the minimum measured scattering vector (assumed to be beyond the small-angle scattering regime), $\left(\mathrm{d}\sigma/\mathrm{d}\Omega\right)_{\mathrm{mag}}$ is the magnetic differential scattering cross section, $S$ is the spin quantum number in units of~$\hbar$, $r_0=\frac{\mu _0}{4\pi}\frac{e^2}{m_e}$ is the classical electron radius, $\gamma = 1.913$ is the neutron magnetic moment in units of nuclear magnetons, $f(Q)$ is the magnetic form factor, $N_s$ is the number of spins in the system, $i$ and $j$ label individual spins~$ \mathbf{S_{\textit i}}$ and~$\mathbf{S_{\textit j}}$ separated by a distance~$r_{ij}$, $A_{ij}=\langle S^y_i S^y_j \rangle$, $B_{ij}=2\langle S^x_i S^x_j \rangle - \langle S^y_i S^y_j \rangle$, $\Theta$ denotes the Heaviside step function, $\rho_0$ is the number of spins per unit volume, and $m$ is the average magnetic moment in \muB\ (zero for anything with no net magnetization, e.g. antiferromagnets). The coordinate system used for~$A_{ij}$ and~$B_{ij}$ for each spin pair is given by $\hat{\boldsymbol{x}}=\frac{\boldsymbol{r_{\textit j}}-\boldsymbol{r_{\textit i}}}{\vert \boldsymbol{r_{\textit j}}-\boldsymbol{r_{\textit i}} \vert}$ and $\hat{\boldsymbol{y}}=\frac{\boldsymbol{S_{\textit i}}-\hat{\boldsymbol{x}}(\boldsymbol{S_{\textit i}}\cdot \hat{\boldsymbol{x}})}{\vert\boldsymbol{S_{\textit i}}-\hat{\boldsymbol{x}}(\boldsymbol{S_{\textit i}}\cdot \hat{\boldsymbol{x}})\vert}$.

In the more general situation where multiple magnetic species and/or orbital contributions to the magnetic moment are present, we use the total angular momentum quantum number $J$ and Land\'e splitting factor $g$ (which need not be the same for each magnetic moment) to express the theoretical mPDF as~\cite{frand;aca14}
\begin{align}
G_{\mathrm{mag}}(r)&=\frac{3}{2\langle g\sqrt{J(J+1)} \rangle^2}\left(\frac{1}{N_s}\sum\limits_{i \ne j}g_i g_j\left[ \frac{A_{ij}}{r}\delta (r-r_{ij})+B_{ij}\frac{r}{r_{ij}^3}\Theta (r_{ij}-r)\right] - 4\pi r \rho_0 \frac{2}{3} m ^2\right) \label{generalfofr},
\end{align}
where $\langle \cdot \cdot \cdot \rangle$ represents an average over all the magnetic moments in the system.

We note that the normalization of $\left(\mathrm{d}\sigma/\mathrm{d}\Omega\right)_{\mathrm{mag}}$ by the square of the magnetic form factor shown in Eq.~\ref{FT} is often difficult to do accurately, particularly if the experimental data contain both nuclear and magnetic scattering. If the Fourier transform is taken without first normalizing the magnetic scattering by the squared form factor, the resulting quantity (often called the unnormalized mPDF) is given by~\cite{frand;aca15}
\begin{align}
d_{\mathrm{mag}}(r)=C_1 \times G_{\mathrm{mag}}(r)\ast S(r) + C_2 \times \frac{\textrm{d}S}{\textrm{d}r},\label{eq;dr}
\end{align}
where $C_1=\frac{N_s}{2\pi}\left(\frac{\gamma r_0}{2}\right)^2 \frac{2}{3}\langle g\sqrt{J(J+1)}\rangle^2$ and $C_2=\frac{N_s}{2\pi}\left(\frac{\gamma r_0}{2}\right)^2 \frac{2}{3}\langle g^2J(J+1)\rangle$ are constants, $\ast$ represents the convolution operation, and $S(r)=\mathcal{F}\left\{f_{m}(Q)\right\}\ast \mathcal{F}\left\{ f_{m}(Q)\right\}$ (where $\mathcal{F}$ denotes the Fourier transform). The quantity $\mathcal{F}\left\{f_{m}(Q)\right\}$ is closely related to the real-space magnetic moment density. The second term on the right side of the equation is unrelated to any pairwise correlations between magnetic moments and results only in a peak at low $r$ (below approximately 1~\AA). Thus, $d_{\mathrm{mag}}(r)$ can be thought of as the properly normalized mPDF $G_{\mathrm{mag}}(r)$ broadened out by approximately $\sqrt{2}$ times the real-space extent of the wave function of the electron(s) giving rise to the magnetic moment. As a final note, when a typical neutron total scattering experiment is performed on a magnetic material and standard data reduction protocols to produce the familiar reduced atomic PDF $G(r)$ are followed, the magnetic component in the total PDF is given by $d_{\mathrm{mag}}(r)/(N_a \langle b \rangle^2)$, where $N_a$ is the total number of atoms in the sample and $\langle b \rangle$ is the average nuclear scattering length.

\subsection{\deltampdf}
The \deltampdf{} is a measure of correlated magnetic disorder in a single crystal~\cite{roth;iucrj18}, analogous to the 3D-$\Delta$PDF for correlated atomic disorder~\cite{weber;zk12}. The \deltampdf\ is obtained via the Fourier transform of the diffuse magnetic scattering, thereby differing from standard PDF techniques where both Bragg and diffuse scattering are included. As such, the \deltampdf\ is sensitive specifically to deviations from the average magnetic structure. In a system without long-range magnetic order, such as a magnetic material above the ordering temperature, there is no average magnetic structure, so the \deltampdf\ reflects all magnetic correlations present. Compared to powder samples, single crystals offer a more feasible route to separating the diffuse scattering from Bragg scattering through methods such as punch and fill~\cite{weber;zk12}, KAREN~\cite{weng;jac20}, or temperature subtraction.

For a given magnetic structure, the \deltampdf{} is given by \cite{roth;iucrj18}
\begin{align}
    \mathrm{3D{\text -}\Delta mPDF}(\mathbf{r}) &= \mathcal{F}^{-1}\left[\frac{d\sigma_{\mathrm{Diffuse}}}{d\Omega} \right]\\
    &=\frac{r^2_0}{4\mu_B^2}\langle \delta \textbf{M} \bar{\otimes} \delta\textbf{M} - \frac{1}{\pi^4} (\delta\textbf{M}\bar{\ast}\boldsymbol\Upsilon)\otimes (\delta\textbf{M}\bar{\ast}\boldsymbol\Upsilon) \rangle \label{eq:3Dmpdf},
\end{align}
where the independent variable $\mathbf{r}$ is the vector separating two points in real space (e.g. the separation vector between two magnetic moments),  $\delta\textbf{M}$ is the vector field describing the local deviations from the average magnetic structure as a function of $\mathbf{r}$, $r_0$ is the classical electron radius as before, and $\mu_B$ is the Bohr magneton. $\boldsymbol\Upsilon$ is a function of $\mathbf{r}$ and arises from the vector interaction of the neutron spin and the electronic magnetic moments in the material. It is defined as 
\begin{equation}
    \boldsymbol\Upsilon \equiv \left\{ \begin{array}{ll}
         \frac{\mathbf{r}}{|r|^4}, & r \neq 0 \\
         \mathbf{0}, & r = 0
    \end{array} \right..
\end{equation}
The vector-field cross correlation and convolution operators ($\bar{\otimes}$ and $\bar{\ast}$) are defined for two vector fields $\mathbf{f}$ and $\mathbf{g}$ as 
\begin{align}
    \mathbf{f}\bar{\otimes}\mathbf{g} &\equiv f_1\otimes g_1 + f_2\otimes g_2 + f_3\otimes g_3 \\
    \mathbf{f}\bar{\ast}\mathbf{g} &\equiv f_1\ast g_1 + f_2\ast g_2 + f_3\ast g_3,
\end{align}
where $f_i$, $g_i$ are the scalar functions comprising the $i^{\mathrm{th}}$ component of the corresponding vector field and $\otimes$ and $\ast$ are the usual scalar cross-correlation and convolution operators, respectively. It can be seen from (\ref{eq:3Dmpdf}) that the \deltampdf{} can be thought of as an auto-correlation of the local magnetization with an additional term to account for the complexity that arises from the neutrons scattering from a vector magnetization density instead of a scalar density. 

\section{Representing magnetic structures in \texttt{diffpy.mpdf}}
At the most basic level, magnetic structures in \texttt{diffpy.mpdf} are represented by arrays containing the magnetic moment vectors and their positions in a sufficiently large volume to calculate the mPDF to the desired distance. Additional information relevant to the magnetic structures, such as the spin and orbital angular momentum quantum numbers and magnetic form factor, can also be supplied. The magnetic moments and their positions can be automatically generated by loading in a magnetic crystallographic information file such as those stored on the MAGNDATA database~\cite{galle;jac16,galle;jac16b}, or users can load in a standard crystallographic information file (or any other structure type accepted by \texttt{diffpy.structure}) and supply appropriate magnetic propagation vectors and basis vectors that describe the magnetic structure. The program accepts both commensurate and incommensurate structures. Alternatively, users can define a magnetic unit cell manually or provide the positions and moments directly. An isotropic atomic displacement parameter (ADP) can also be defined for the magnetic structure. The value of this parameter makes a significant difference for the normalized mPDF, but realistic ADP values have a negligible effect on the unnormalized mPDF due to the much more significant broadening arising from the magnetic form factor.

To enable effective modeling of short-range magnetic correlations, \diffpympdf\ also allows users to define a finite correlation length for any given magnetic structure. In the simplest case, an isotropic correlation length $\xi$ is used such that the magnitudes of the magnetic moments decrease as $\exp\left(-r/\xi\right)$ with distance $r$ from the magnetic moment taken to be at the origin in the calculation. The effect of this is almost identical to multiplying the ideal mPDF pattern assumed to have an infinite correlation length by the exponential envelope $\exp\left(-r/\xi\right)$. Because the correlations may not always decay isotropically, we also introduce in \diffpympdf\ a new way to model anisotropic correlations in which the correlation length along any given direction is represented by the surface of an ellipsoid. Details are provided in the Appendix. An anisotropic correlation length of this type was recently used to describe the short-range magnetic correlations in the antiferromagnetic semiconductor MnTe~\cite{baral;matter22}.

\section{1D mPDF analysis}
\subsection{Calculating the mPDF}
The mPDF for a given magnetic structure can be calculated for any user-specified $r$ range and step size. The program can calculate both the normalized and unnormalized mPDF patterns, as desired by the user. The calculations also include the effects of the experimental parameters $Q_{\mathrm{max}}$, $Q_{\mathrm{min}}$, and $Q_{\mathrm{damp}}$ if provided by the user. 

\subsection{Additional capabilities}
The \diffpympdf\ package has several other built-in capabilities useful for or related to 1D mPDF analysis. These include extracting the magnitude of the locally ordered magnetic moment from combined atomic and magnetic PDF fits, applying a Fourier filter to data to remove components that cannot physically be magnetic in origin, calculating the magnetic scattering pattern from an mPDF pattern via the Fourier transform, and subtracting a high-temperature PDF from a low-temperature PDF to isolate the magnetic component with built-in optimization to reduce artifacts introduced by thermal effects.

\subsection{Examples}
\begin{figure}
\includegraphics[width=10.0cm]{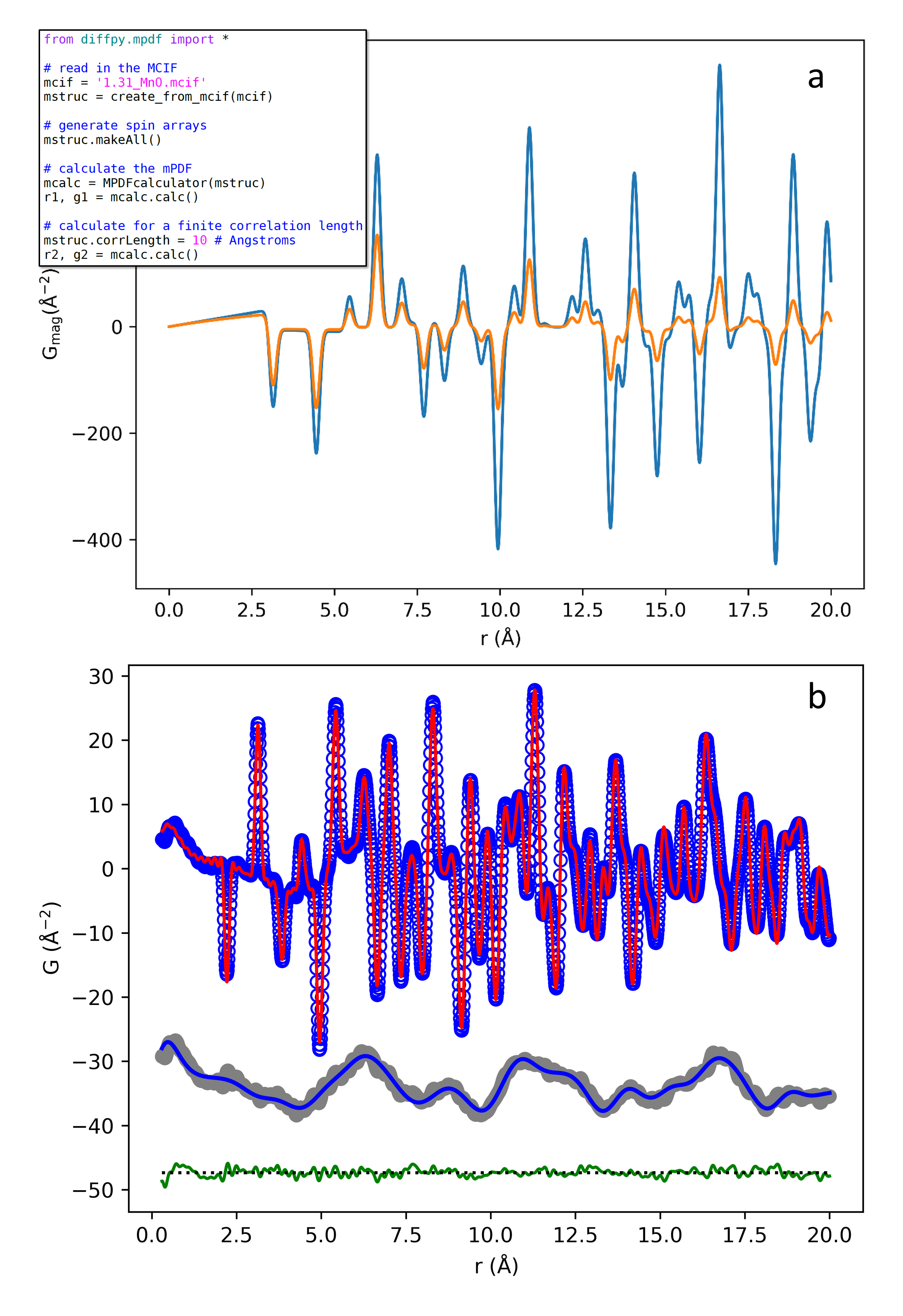}
\caption{(a) Calculated mPDF patterns for MnO assuming an infinite correlation length (blue curve) and a finite correlation length of 10~\AA\ (orange curve). Inset: Python code used to generate the mPDF patterns. (b) Combined atomic and magnetic PDF fit for MnO at 5~K. The upper set of curves displays the total PDF data (blue circles) and calculation (red curve); offset just below is the isolated mPDF (gray curve) and calculated mPDF (blue curve); and the overall fit residual is given by the green curve near the bottom.}
\label{fig:MnO}
\end{figure}

We provide two examples of 1D mPDF calculations and fits using the antiferromagnetic insulator MnO and the ferromagnetic semimetal MnSb. In Fig.~\ref{fig:MnO}(a), we calculate the normalized mPDF for MnO assuming an infinite correlation length (blue curve) and an isotropic correlation length of 10~\AA\ (orange curve), resulting in a pronounced damping effect in real space. 
The inset in Fig.~\ref{fig:MnO}(a) displays the python code snippet used to calculate the two mPDF curves. Fig.~\ref{fig:MnO}(b) displays a combined atomic and magnetic PDF fit to MnO at 5~K (data collected on the NOMAD instrument~\cite{neuef;nimb12} at the Spallation Neutron Source following a similar procedure as in~\cite{frand;aca15}). The blue circles show the experimental data, with the best-fit total PDF (atomic plus magnetic) overlaid in red. Offset vertically below, we display the isolated mPDF component in gray (assumed to be the total PDF signal with the best-fit atomic PDF component subtracted out, which ideally yields the unnormalized mPDF), with the best-fit unnormalized mPDF shown by the thick blue curve. The overall fit residual is given by the green curve. The broad peak below about 1~\AA\ corresponds to the second term on the right hand side of Eq.~\ref{eq;dr}. The ordered moment determined from the fit is 4.9~$\mu_{\mathrm{B}}$, consistent with expectations for the $S = 5/2$ Mn$^{2+}$ spins. We use the unnormalized mPDF for the fit because the scattered intensity is not normalized by the square of the magnetic form factor in the standard PDF data reduction protocols used here, resulting in a combined total PDF signal consisting of the proper nuclear PDF and the unnormalized mPDF. Naturally, the properly normalized mPDF can subsequently be calculated from the refined model of the magnetic structure if desired.

\begin{figure}
\includegraphics[width=10.0cm]{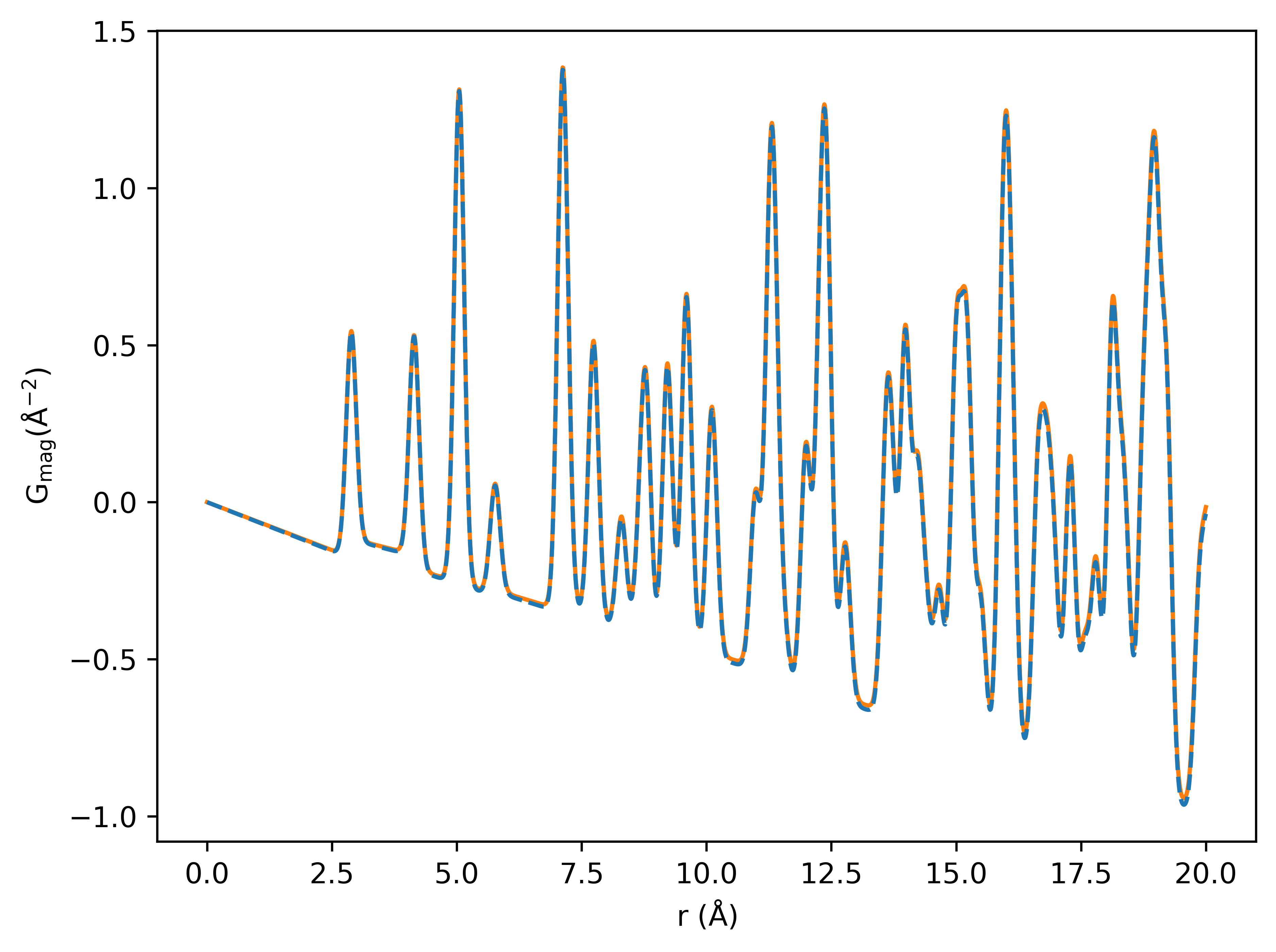}
\caption{Calculated mPDF pattern for ferromagnetic MnSb. The dashed blue curve was calculated using the exact formula with the density and magnetization, and the solid orange curve was calculated after automatically estimating the slope of the linear term, as described in the main text. }
\label{fig:MnSb}
\end{figure}
The calculated mPDF for ferromagnetic MnSb is shown in Fig.~\ref{fig:MnSb}. In this case, the mPDF contains only positive peaks, since all the magnetic moments are aligned parallel to each other. As such, the linear term in Eq.~\ref{fullfofr} proportional to $m^2$ is nonzero. The slope of the linear term can be calculated exactly using the average density $\rho_0$ and magnetization $m$ per magnetic moment, or it can be determined empirically in \texttt{diffpy.mpdf} by requiring the average value of $G_{\mathrm{mag}}$ to be zero. These two methods are shown by the dashed blue and solid orange curves, respectively, in Fig.~\ref{fig:MnSb}. The difference between them is negligible in this case.

\section{\deltampdf\ analysis}
\subsection{Calculating the \deltampdf}

The \deltampdf{} is calculated by evaluating Eq.~\ref{eq:3Dmpdf} for a given set of magnetic moment vectors and position vectors representing the magnetic structure. We convolve each position vector with a user-specifiable 3D Gaussian function, approximating the effect of the finite spatial extent of the wave functions of the unpaired electron(s) giving rise to each magnetic moment. The user can also specify the 3D real-space grid on which the \deltampdf\ is to be evaluated.
\subsection{Example}
We present the \deltampdf{} of the antiferromagnetic semiconductor MnTe as an example case. MnTe has a hexagonal crystal structure and orders magnetically below $T_{\mathrm{N}}=307$~K, with the moments aligned ferromagnetically within the \textit{ab} plane and antiferromagnetically along the \textit{c} axis~\cite{kunit;jdp64}. Short-range antiferromagnetic correlations are known to persist to high temperatures well into the paramagnetic phase~\cite{zheng;sadv19,baral;matter22}.
\begin{figure}
    \centering
    \includegraphics[width=10.0cm]{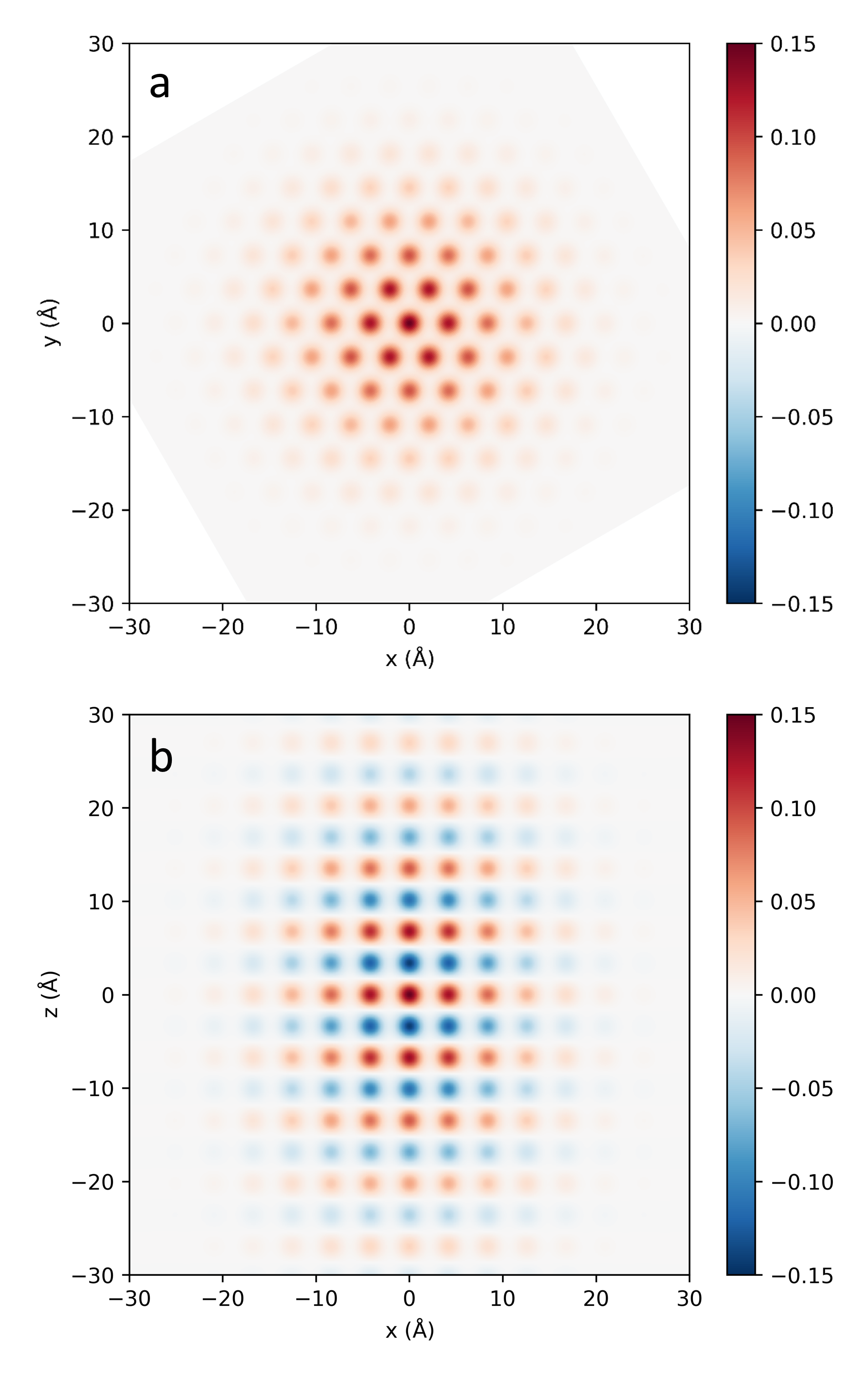}
    \caption{(a) The $z=0$ plane of the calculated \deltampdf{} pattern for MnTe above \TN. The calculation shows short-range ferromagnetic correlations in the \textit{xy} plane. (b) The $y=0$ plane of the calculated \deltampdf{} pattern for MnTe above \TN. The calculation shows short-range antiferromagnetic correlations in the \textit{z} direction. The units shown in the color bar for both panels are arbitrary, but the sign is significant: for a given separation vector, a positive (negative) value of the \deltampdf{} indicates greater parallel (antiparallel) alignment of the magnetization relative to the average magnetic structure. In the present case of a correlated paramagnet, there is no average magnetic structure, so the \deltampdf{} shows the local magnetic correlations directly.}
    \label{fig:MnTe_xy}
\end{figure}

In Fig.~\ref{fig:MnTe_xy}(a), we show the $z=0$ plane of the calculated \deltampdf{} for MnTe in the correlated paramagnet regime above \TN. The magnetic configuration was generated using experimentally verified \textit{ab initio} calculations~\cite{baral;matter22}. The hexagonal structure and ferromagnetic alignment within the plane can both be seen. In Fig.~\ref{fig:MnTe_xy}(b), we show the $y=0$ slice of the same calculation, providing a view of the out-of-plane correlations. The horizontal rows of alternating red and blue spots illustrate the antiferromagnetic correlations along the \textit{z} axis. We also note the strikingly anisotropic correlation length, which is significantly longer along \textit{z} than within the \textit{xy} plane. This originates from the disparate strengths of the magnetic exchange interactions~\cite{baral;matter22}. 

\section{Availability and resources}
The \texttt{diffpy.mpdf} package is open source and distributed under the 3-Clause BSD license. The source code is freely available to download and install at https://github.com/FrandsenGroup/diffpy.mpdf. The software runs on Linux and Macintosh operating systems and the Windows Subsystem for Linux. Tutorial files in the form of Jupyter Notebooks and python scripts are also available at the same website to demonstrate important functionalities, including creating magnetic structures, calculating one- and three-dimensional mPDF patterns, performing fits to data, and a selection of more advanced usage cases. 

\appendix

\section{Ellipsoidal model of an anisotropic correlation length}
We represent an anistropic correlation length using an ellipsoid, where the distance from the center to the surface of the ellipsoid in any given direction is the correlation length in that direction. The surface of an ellipsoid is described by the equation
\begin{equation}\label{eq:ellipsoid}
    x^TDx=1,
\end{equation}
where $x$ is the three-component Cartesian vector from the center of the ellipsoid to an arbitrary point on the surface of the ellipsoid and $D$ is a symmetric matrix of the form 
\begin{equation}
    D = \begin{bmatrix}
    D_{11} & \frac{1}{2} D_{12} & \frac{1}{2} D_{13} \\
    \frac{1}{2} D_{12} & D_{22} & \frac{1}{2} D_{23} \\
    \frac{1}{2} D_{13} & \frac{1}{2} D_{23} & D_{33}
    \end{bmatrix}.
\end{equation}
If we consider an arbitrary direction denoted by the unit vector $\hat{n}$, we can determine the correlation length $\xi$ along that direction by substituting the vector $\xi\hat{n}$ for $x$ in Eq.~\ref{eq:ellipsoid}, yielding the desired equation for $\xi$
\begin{equation}
    \xi = (\hat{n}^TD\hat{n})^{-\frac{1}{2}}.
\end{equation}
Thus, one can implement any desired ellipsoidal correlation length by constructing a suitable $D$ matrix. Because the components of $D$ are related to the rate at which magnetic correlations decay along certain directions, we call $D$ the damping matrix.

The eigenvectors of $D$ represent the principal axes of the correlation ellipsoid, while the eigenvalues are the inverse square of the correlation length along each of the principal axes. If physical considerations suggest a certain set of principal axes for the correlation ellipsoid, then we can use these principal axes and their associated correlation lengths to construct $D$ through spectral decomposition:
\begin{align}
    D &= V\Lambda V^T \\
    &= [n_1,n_2,n_3] \begin{bmatrix}
    \frac{1}{\xi_1^2}&0&0 \\
    0&\frac{1}{\xi_2^2}&0 \\
    0&0&\frac{1}{\xi_3^2}
    \end{bmatrix} 
    \begin{bmatrix}
        n_1^T \\
        n_2^T \\
        n_3^T \\
    \end{bmatrix}.
\end{align}
The \diffpympdf\ package allows users to construct the damping matrix directly or by providing a set of orthornormal eigenvectors (i.e., the directions of the principal axes of the ellipsoid) and their correlation lengths. 

\ack{Acknowledgements}

We thank Caleb Dame for valuable contributions to the interactive magnetic structure builder and Matthew Richards for help with the fitting algorithms. Work by B.A.F. and P.H. was supported by the U.S. Department of Energy, Office of Science, Office of Basic Energy Sciences (DOE-BES) through Award No. DE-SC0021134. J.A.C. and E.S. acknowledge support from the College of Physical and Mathematical Sciences at Brigham Young University. S.J.L.B. was supported by U.S. Department of Energy, Office of Science, Office of Basic Energy Sciences (DOE-BES) under contract No. DE-SC0012704. This study used resources at the Spallation Neutron Source (SNS), a DOE Office of Science User Facility operated by the Oak Ridge National Laboratory.

\end{document}